\documentclass[draft,grl]{AGUTeX}

 \usepackage{lineno}
 \linenumbers*[1]

\usepackage{graphicx}
\def\grl{{\em Geophys. Res. Lett.}}
\def\jgr{{\em J. Geophys. Res.}}
\def\prl{{\em Phys. Rev. Lett.}}

\def\pr{{\em Physical Review }}

\def\sci{{\em Science}}
\def\pop{{\em Phys. Plasma}}

\authorrunninghead{Che ET AL.}

\titlerunninghead{LH Instability and Electron Holes}

\authoraddr{Che, H., J. F. Drake, and M. Swisdak,
IREAP, Department of Physics, University of Maryland, College Park, MD, USA }
\authoraddr{J. F. Drake, and P. H. Yoon,
IPST, University of Maryland, College Park, USA }

\begin{document}
\title{Electron Holes and Heating in the Reconnection Dissipation Region}
\authors{H. Che, \altaffilmark{1},
J. F. Drake, M. Swisdak  and  P. H. Yoon }


\altaffiltext{1}{Current Address: Center for Integrated Plasma  Studies (CIPS), Department of Physics, University of Colorado, UCB 390, Boulder, CO, USA }
%
%


\begin{abstract}
Using particle-in-cell simulations and kinetic theory, we explore the
current-driven turbulence and associated electron heating in the
dissipation region during 3D magnetic reconnection with a guide
field. At late time the turbulence is dominated by the Buneman and
lower hybrid instabilities. Both produce electron holes that co-exist
but have very different propagation speeds. The associated scattering of
electrons by the holes enhances electron heating in the dissipation region.

\end{abstract}

\begin{article}

%
%

\section{Introduction}
Magnetic reconnection is the driver of explosive events in nature,
such as solar flares, substorms in the magnetosphere of the Earth and
flares from magnetars and the accretion disks
of black holes. Satellite observations in the Earth's magnetosphere
indicate that magnetic reconnection drives turbulence. Electron holes,
which are localized, positive-potential structures caused by plasma
kinetic instabilities, have been linked to current sheets associated
with magnetic reconnection in the magnetotail
\citep{farrell02grl,cattell05jgr,andersson09prl}, the
magnetopause\citep{matsumoto03grl}, and the
laboratory\citep{fox08prl}. Lower hybrid (LH) waves and other plasma
waves appear in conjunction with electron holes in the magnetotail
events. Electron holes can scatter electrons, causing heating and
possibly anomalous resistivity to facilitate fast magnetic
reconnection.

During magnetic reconnection, a parallel electric field generated
around the x-line drives electron beams. Simulations with a guide
field show that these intense beams can drive the Buneman instability,
which forms bipolar structures in the parallel electric field
\citep{drake03sci}. Later in time transverse electric fields
develop. Following a suggestion that these transverse fields were
current-driven lower hybrid waves (LHI) \citep{mcmillan06pop}, Che et
al. \citep{che09prl} showed that both the LH and electron-electron
two-stream instabilities resonate with the high velocity electrons and
therefore dominate the interactions with the highest velocity
electrons in narrow current layers. Which instabilities develop during
reconnection and how they interact remains unknown.

During magnetic reconnection we demonstrate that two distinct classes
of electron holes with very different propagation speeds exist
simultaneously. Slow moving holes are driven by the Buneman
instability and at the same time and locations fast moving holes are
driven by the LHI. Both take the form of nonlinear
Bernstein-Greene-Kruskal (BGK) solutions \citep{bernstein57pr} since
the measured bounce time of electrons in the holes is short compared
with the hole lifetime. The trapping and scattering of electrons by
holes of disparate phase speed enhances dissipation during
reconnection.

\section{Simulation }

We carry out 3D magnetic reconnection simulations with a strong guide
field similar to those carried out earlier \citep{drake03sci} but with
a much larger simulation domain: $L_x= 4 d_i$, $L_y=2 d_i$, and $L_z=4
d_i$, where $d_i= c/\omega_{pi}$ and $\omega_{pj}$ is the plasma
frequency of a particle species $j$.  The reconnecting magnetic field
is $B_x/B_0=\tanh[(y-L_y/4)/w_0]-\tanh[(y-3L_y/4)/w_0]-1$, where $B_0$
is the asymptotic amplitude of $B_x$ outside of the current layer, and
$w_0$ is the half-width of the initial current sheet. The guide field
$B_z^2 = B^2-B_x^2$ is chosen so that the total field $B$ is
constant. In our simulation, $B$ is taken as $26^{1/2} B_0$.  The
initial temperature is $T_e =T_i =0.04 m_i c_A^2$, the ion to electron
mass ratio is $100$, the speed of light $c$ is $20 c_A $ with $c_A=
B_0/(4 \pi n_0 m_i)^{1/2}$, the Alfv\'en speed. The initial drift
speed of $4c_A$ is just above the electron thermal speed $3c_A$ and
marginally exceeds the threshold to trigger the Buneman instability.

\begin{figure}
\includegraphics[scale=0.65, trim=0 5 100 0,clip]{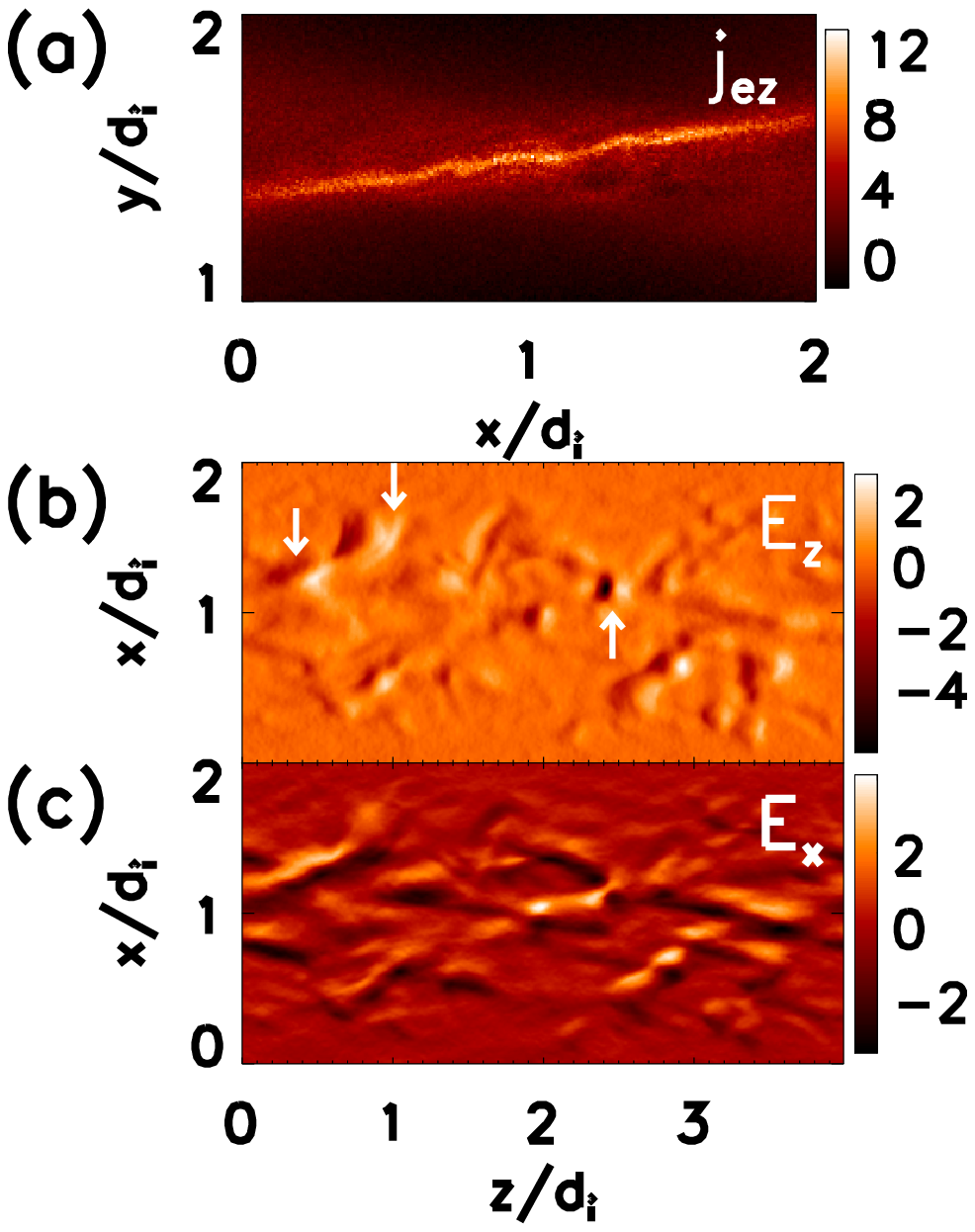} 
\caption{{\bf (a)}: The current sheet $j_{ez}$ in the $x-y$ plane at $\Omega_i t =3.3$. {\bf (b, c)}: The spatial structures of the electric fields $E_x$ and $E_z$ in the $x-z$ plane in a cut through the current layer.}
\label{lhifig1}
\end{figure}

\begin{figure}
\includegraphics[scale=0.35, trim=10 430 0 20,clip]{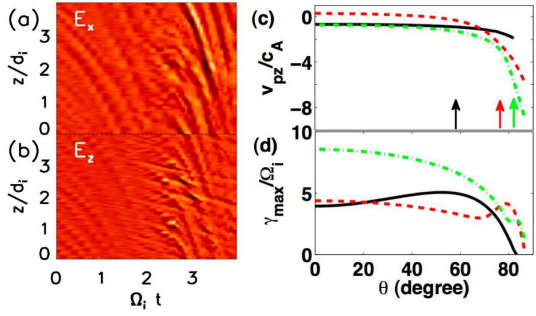} 
\caption{ {\bf (a, b)}: Cuts of $E_x(z)$ and $E_z(z)$ around the
x-line at different times from the simulation. {\bf (c)}: The
theoretical parallel phase speed $v_{pz}$ vs. the angle $\theta$
between wavevector $\mathbf{k}$ and magnetic field at $\Omega_i
t=1,3,4$ (black solid, red dashed and green dash-dotted lines). The
arrows denote the angle $\theta$ of the fastest-growing mode of the LH
instability at the three times in (d). {\bf (d)} The theoretical growth rate $\gamma_{max}$ of fast-growing mode vs. the angle $\theta$ at the three times in (c). }
\label{lhifig2}
\end{figure}

\begin{figure}
\includegraphics[scale=0.5, trim=0 10 0 0,clip]{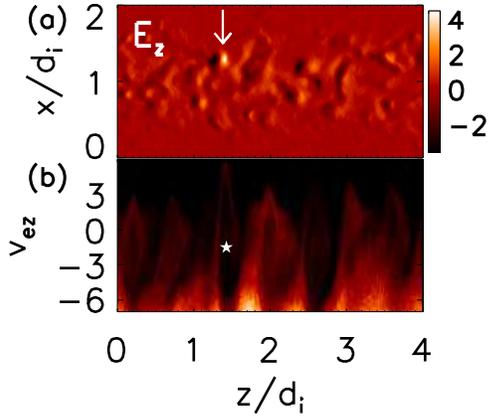} 
\caption{{\bf (a)}: Spatial structure of $E_z$ at $\Omega_i t =3$ in
the current layer. {\bf (b)}: The phase space $z-v_{ez}$ at $x\sim
1.2$ of (a).}
\label{lhiez}
\end{figure}

Magnetic reconnection induces a parallel electric field around the
x-line and drives an intense electron beam. At $\Omega_i t =3 $
($\Omega_i = eB_0/m_i c$), the electron beams have been accelerated to
$10 c_A$ and to $14 c_A$ at $\Omega_i t =4 $. We show the current
sheet around the x-line in the $x-y$ plane at $\Omega_i t =3.3$ in
Fig.~\ref{lhifig1}~(a). At the beginning of the magnetic reconnection
simulation, the Buneman instability with wavevector along the magnetic
field $z$ direction is excited. In the cold plasma limit, the phase
speed is $(m_e/(2 m_i))^{1/3}|v_{dz}|/2 \sim 1 c_A $ and the growth
rate is $\gamma \sim \sqrt{3}\omega_{pe}(m_e/(2 m_i))^{1/3}/2\sim 29
\Omega_i $ \citep{galeev84book1}. The Buneman instability saturates
within a short time. Later in time two distinct spatial structures of
the electric field are observed: localized bipolar structures dominate
$E_z$ and long oblique stripes dominate $E_x$. A surprise is that
there are two types of bipolar structures. At $\Omega_i t=3$ one has a
velocity close to zero and the other moves with a velocity of $3
c_A$. By $\Omega_i t=4$ the velocity of the second increases to $7
c_A$. In Fig.~\ref{lhifig1}~(b, c) we show $E_z$ and $E_x$ in the
midplane $x-z$ of the current sheet at $\Omega_i t =3.3$. The
structures move to the left in this figure, which is in the direction
of the electron drift. The downward (upward) arrows point to fast
(slow) moving electron holes.  To see the two classes of holes more
clearly, in Fig.~\ref{lhifig2}~(a, b) we stack cuts of $E_x(z)$ and
$E_z(z)$ at the x-line versus time. The dark and light bands mark the
development of the bipolar structures seen in Fig.~\ref{lhifig1}~(b,
c). The slopes of these bands are the phase speeds of the
waves. During the time interval $\Omega_i t = 0-2$, the phase speed of
the waves increases, which was expected since the streaming velocity
of the electrons increased as the reconnection driven current layer
shown in Fig.~\ref{lhifig1}~(a) developed. During the time interval
$\Omega_i t =2-4$ two distinct phase speeds, particularly in $E_z$,
are evident. In Fig.~\ref{lhifig2}~(b) the structures cross each other
at the same value of $z$, which indicates that this result is not due
to the spatial structure of the streaming velocity.  In
Fig.~\ref{lhiez} we show $E_z$ and the $z-v_z$ phase space around
$(x,y)=(1.2d_i,1.5d_i)$ at $\Omega_it=3$ to reveal the structure of the
fast moving holes. There are no slow holes in this region at this
time. In (a) the most intense hole is marked by the arrow. In (b) the
center of the $z-v_{ez}$ phase space of this bipolar structure is
marked by the star. The electrons encircling the star indicate that
electrons are trapped by the bipolar field. The strong electron
heating due trapping is evident. 

Electron holes in the simulation exhibit a complex dynmics: formation,
dissipation and reformation. The lifetimes $\tau_l$ of the two classes
of electron holes are distinct, around $0.1 \Omega_i^{-1}$ and $0.2
\Omega_i^{-1}$ for the fast and slow holes, respectively. In both
$\tau_l$ exceeds the bounce time of the trapped electrons, $\tau_{b}
\approx \sqrt{m_e\lambda_b}/\sqrt{2e \delta E_z} \sim 0.02
\Omega^{-1}_i$, where $\lambda_b$ is the characteristic wavelength of
the electron hole. Thus, electron trapping takes place and we
therefore interpret the holes as BGK
structures. \citep{bernstein57pr}.

\section{Kinetic Model and Analytic Results }

We now investigate which instabilities drive the two distinct types of
holes by examining in more detail the development of streaming
instabilities. Using two drifting Maxwellians to model the electron
distribution and a single Maxwellian to model the ion distribution, we
fit the distribution functions obtained from the simulations and
substitute the theoretical fittings into the local dispersion function
derived from kinetic theory for waves with $\Omega_i \ll \omega \ll
\Omega_e$ \citep{chephd}:
\begin{eqnarray}
1+\frac{2\omega_{pi}^2}{k^2 v^2_{ti}}[1+\zeta_i Z(\zeta_i)] 
+ \frac{2 (1-\delta)\omega^2_{pe}}{k^2 v^2_{te1}}[1+I_0(\lambda) e^{-\lambda} \zeta_{e1} Z(\zeta_{e1})]\\ \nonumber
+ \frac{2 \delta\omega^2_{pe}}{k^2 v^2_{te2}}[1+I_0(\lambda) e^{-\lambda} \zeta_{e2} Z(\zeta_{e2})]=0,
\label{bdf}
\end{eqnarray}
where $\zeta_i=(\omega-k_{z} v_{di})/k v_{ti}$, $\zeta_{e1} =
(\omega-k_{z} v_{de1})/k_{z} v_{z te1}$, $\zeta_{e2} = (\omega-k_{z}
v_{de 2})/k_{z} v_{z te2}$, $\lambda=k^2_x v^2_{x te}/2\Omega^2_e$,
$\delta$ is the weight of the low velocity drifting Maxwellian, $Z$ is
the plasma dispersion function and $I_0$ is the modified Bessel
function of the first kind with order zero. The thermal velocity of
species $j$ is defined by $v^2_{tj}=2T_{tj}/m_j$ and drift speed by
$v_{dj}$, which is parallel to the magnetic field ($z$ direction). The
electron temperature takes a different value along and across the
magnetic field while the ions are taken to be isotropic.

\begin{figure}
\includegraphics[scale=0.4, trim=30 380 60 40,clip]{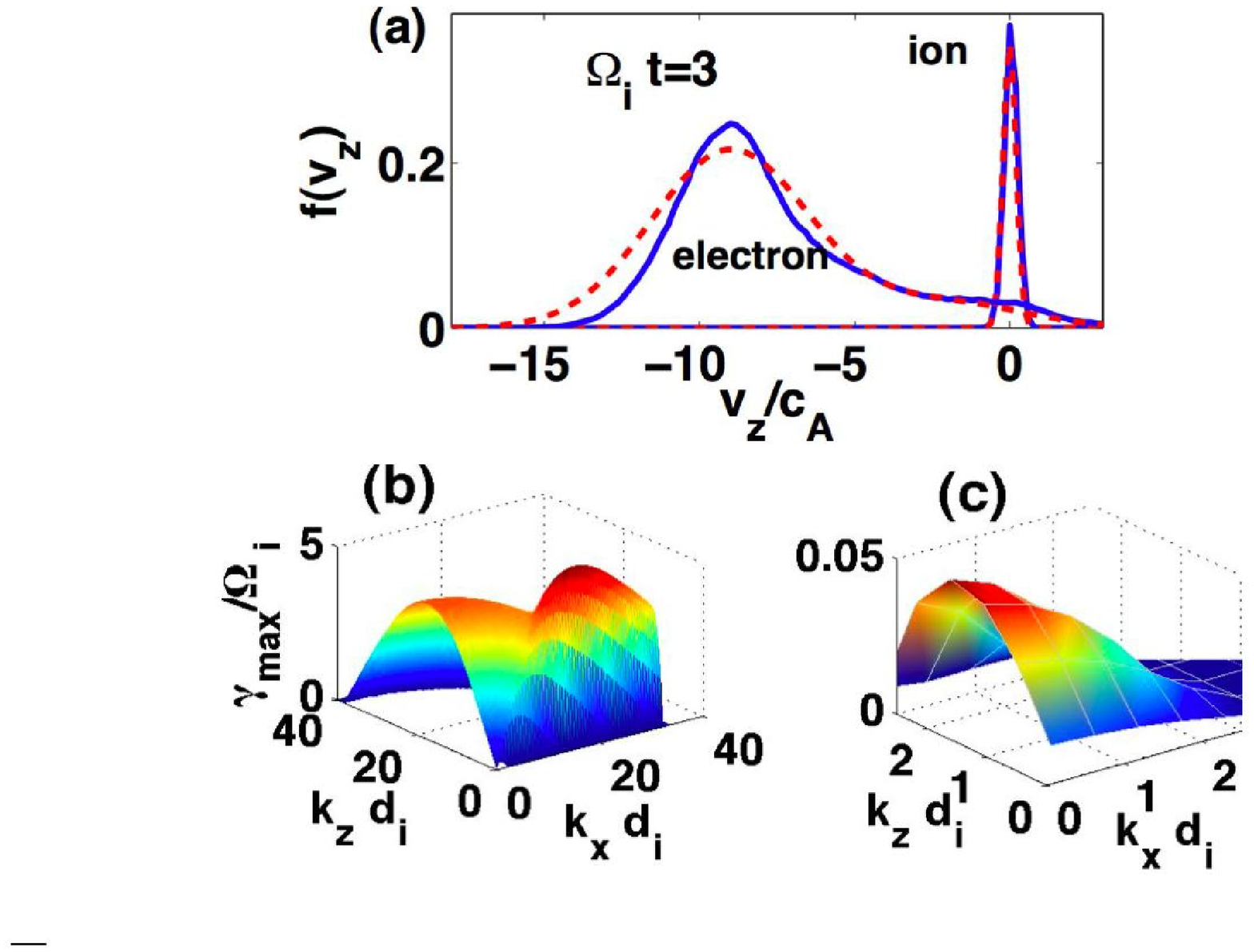} 
\caption{{\bf (a)}: Electron and ion distribution functions $f(v_z)$
around the x-line at $\Omega_i t = 3$ from simulations (blue solid) and the model (red-dashed) with the ion distribution function reduced by a factor of
four. In (b) the 2D spectrum includes both
electrons and ions and in (c) is without the ions.}
\label{lhifig3}
\end{figure}

The fitting parameters of the distribution functions at $\Omega_i t
=3,4$ are listed in Table \ref{lhipara}. The match between the
parallel distribution and our fitted distribution is shown in
Fig.~\ref{lhifig3}~(a). We can see from the Table that the weight
$\delta$ of the low velocity electrons increases with time, indicating
that momentum is transferred from the high velocity to the low
velocity electrons.

The theoretical 2D spectrum at $\Omega_i t = 3$ is shown in
Fig.~\ref{lhifig3}~(b). Two distinct modes are found, one with
$\mathbf{k}$ parallel and the other with $\mathbf{k}$ nearly
perpendicular to $\mathbf{B}$. The peak of the parallel mode is around
$ k_z d_i \sim 20$, which is close to the wavenumber of the cold
plasma limit of the Buneman instability, $k_z d_i = \omega_{pe}/v_{de}
\sim 20$. To confirm that the parallel mode is the Buneman instability
rather than the two-stream instability, we exclude ions from our
calculations. The mode obtained only with electrons is shown in
Fig.~\ref{lhifig3}~(c). The two-stream instability has a much smaller
growth rate. Thus, the parallel mode is the Buneman instability. The
peak of the nearly-perpendicular mode is centered at $(k_x d_i, k_z
d_i) = (22,5)$. The frequency of this mode is $ \sim 13\Omega_i$ which
is in the LH frequency range for the present simulation so the
nearly-perpendicular mode is the LHI
\citep{mcmillan06pop,che09prl}. 

\begin{figure}
\includegraphics[scale=0.45, trim=30 570 0 30,clip]{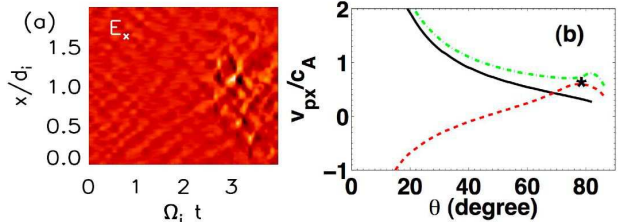} 
\caption{{\bf (a)}: Cuts of $E_x(x)$ at different times from the
simulation. ({\bf b}): Theoretical phase speed $v_{px}$ vs. $\theta$
at $\Omega_i t= 1,3,4$, denoted by black solid, red-dashed and green
dash-dotted lines.}
\label{lhifig4}
\end{figure}

As a test of this interpretation, we compare the phase speed of the
modeled waves across ($v_{px}$) and along ($v_{pz}$) $\mathbf{B}$ with
the simulation data. The assumption here is that since the fraction of
trapped electrons in any given electron hole is small, the non-trapped
particles control the phase speeds of the wave and the linear
dispersion characteristics can be used to interpret hole propagation.
It is well known that the Buneman instability can form parallel
bipolar structures. This instability, which has a very low parallel
phase speed to enable coupling to the ions, is the source of the
electron holes moving slowly parallel to the magnetic field. Thus, the
LHI should be responsible for the oblique, fast-moving electron holes
marked by the downward arrows in $E_z$ and the oblique stripes in
$E_x$ in Fig.~\ref{lhifig1}. This interpretation is consistent with
the parallel phase speeds $v_{pz}$ of the Buneman and LH instabilities
obtained by the kinetic model which are shown in
Fig.~\ref{lhifig2}~(c). The phase speed of the Buneman instability
with $\theta \sim 0$ is close to zero. The three arrows from left to
right (black,red and green) indicate the position $\theta$ of the
maximum-growing mode of the LH instability at $\Omega_i t =1,3,4$ shown in Fig.~\ref{lhifig2}. The
phase speed of the LH instability is initially low and then increases
to $4 c_A$ at $\Omega_i t =3$ and to $7 c_A$ at $\Omega_i t =4$ . The
high phase speed of the LHI is consistent with the fast-moving
electron holes seen at late time in the simulation. As a further check
on this interpretation, in Fig.~\ref{lhifig4} (a) we stack the cuts of
$E_x(x)$ along $x$ at different times. The slope of the curves is the
phase speed $v_{px}$. We see that at $\Omega_i t =3$ $v_{px}\sim 0.6
c_A$. In (b) is the theoretical phase speed $v_{px}$ at $\Omega_i t
=1,3,4$ calculated from the model. At $\Omega_i t =3$ the $v_{px}$ of
the LH wave, marked with the ``*'', is around $0.6 c_A$, consistent
with the value from the simulation.


\section{Conclusion}
In summary, we have demonstrated through simulations and an analytic
model that two distinct classes of electron holes are generated
simultaneously in the intense current layers that form during magnetic
reconnection. The sources of the holes are the Buneman and LHI. The LH
waves produce a transverse field $E_x$ as well as the bipolar
structures $E_z$ that trap electrons to form electron holes. These
electron holes move along the magnetic field at the phase speed of the
LH wave. Electron holes formed by the Buneman instability move more
slowly. The simultaneous existence of electron holes with two distinct
phase speeds enables electron scattering over a much larger range of
velocity space than would be possible by either either instability
alone. Electron dissipation in the intense current layers that form
during reconnection is therefore enhanced. The LH electron hole was
also independently observed by 2D Vlasov simulations
\citep{newman08agu}.

%
%
%
%
%
%

%
%
%
%

\begin{acknowledgments}
This work was supported in part by NSF ATM0613782, and NASA NNX08AV87G and NWG06GH23G. PHY acknowledges NSF grant ATM0837878. HC thanks Drs. M. Goldman and D. Newman for their helpful comments. The simulations were carried out at the National Energy Research Scientific Computing Center.
\end{acknowledgments}


\begin{table}[h]
\caption[Short title]{Parameters of Model Dist. Funs.}
\begin{center}
\begin{tabular}{|c|c|c|c|c|c|c|c|c|}
\hline & $v_{xte}$ & $v_{zte1}$ & $v_{zte2}$ & $v_{de1}$ & $v_{de2}$ & $v_{ti}$  & $v_{di}$  & $\delta$ \\
\hline $\Omega_i t$= 3 & 2.8 & 3.6 & 3.5 & -9.0 & -2.0 & 0.3 & 0 & 0.16 \\
\hline $\Omega_i t$= 4 & 2.8 & 4.0 & 4.2 & -9.0 & -5.0 & 0.34 & 0.1 & 0.26 \\
\hline
\end{tabular}
\end{center}
\label{lhipara}
\end{table}

\end{article}
\end{document}